\documentclass[12pt]{revtex4}

\usepackage{graphicx}
\usepackage{dcolumn}
\usepackage{bm}

\begin{document}

\title[Growth Model for Vote Distributions in Elections]{Growth Model for Vote Distributions in Elections}

\author{Chung-I Chou$^1$, Sai-Ping Li$^{2,3}$}

\affiliation{$^1$Department of Physics, Chinese Culture University, Taipei, Taiwan 111, R.O.C.}
\affiliation{$^2$Institute of Physics, Academia Sinica, Taipei, Taiwan 115, R.O.C.}
\affiliation{$^3$Department of Physics, University of Toronto, Toronto, Ontario M5S 1A7, Canada}

\begin{abstract}
There are many factors that can influence the outcome of an election.  
We here identify two dominant effects that can affect the votes obtained by a candidate, 
namely, the Majority Effect and the Media Effect.  We mimic these two effects  
in a simple growth model.  We put the model on a two-dimensional
square lattice and test it against the available data from elections in various
countries.  By adjusting the two parameters in the model, we are
able to fit the vote distributions in all the countries studied. 
\end{abstract}

\maketitle

\section{Introduction}
It is now a common practice for a democratic country to hold
elections in order to choose an individual to hold formal office.  In elections, 
there are indeed many factors that can influence the outcome of the result.  It is
noted that the outcome of the distribution of votes for individual candidates has 
taken interest from researchers in recent years[1-11].  Different countries however, hold
regular elections by adopting different electoral systems.   In fact, with different 
electoral systems, one would imagine that the vote distributions
of candidates in elections would look very different.  In the case of 
proportional elections held in Brazil\cite{Costa Filho}, the distribution of votes
among candidates for the whole country was shown to follow a power
law $N(x) \propto x^{-\alpha}$  with $\alpha = 1.00 \pm 0.02$. On the
other hand, in many European countries, the vote distributions turn 
out to follow a lognormal distribution\cite{Fortunato}.   Since different
countries adopt different electoral systems, one would probably 
argue that these different distributions are indeed results of different electoral systems.  

Figure 1 is an illustration of the vote distributions of
elections from different countries\cite{ElectionData}.  Figures 1(a)-1(f) are vote distributions 
from Brazil, Finland, USA, Australia, Canada and Taiwan.  The bracket next to the
name of the country is the electoral system that the country is adopting.  
Here, $x$ is the number of votes obtained by the candidate divided by the total 
number of votes in the election and $P(x)$ is the probability density.  
It is interesting to note 
that vote distributions in Fig. 1(c)-(f) display two peak behavior.  This is significantly
different from either power law or lognormal distributions as in Fig. 1(a) and (b).  The 
apparent differences could probably be thought of related to the different electoral
systems used in different countries.  However, in the case
of Taiwan, the country has changed its electoral systems in 2008 but the essential
features of 
the vote distributions in elections before and in that year is basically unchanged.  
Therefore it is natural to ask if different vote
distributions are in fact results coming from the same 
mechanism that governs elections instead of different electoral systems.  

\begin{figure}
\includegraphics[width=1.0\linewidth]{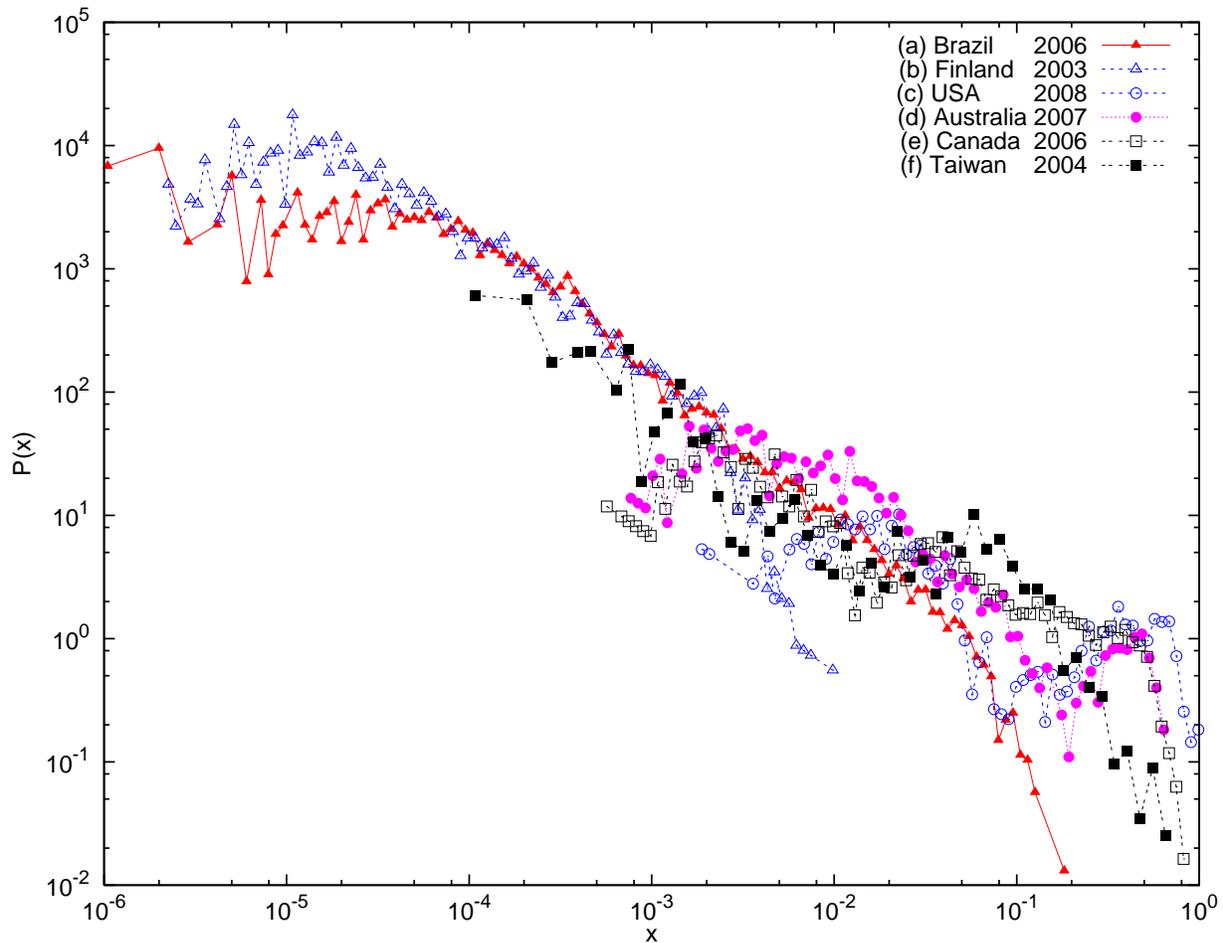}

\caption{\label{fig:1} Vote distributions from (a)Brazil(Proportional representation system, large voting district), (b)Finland(Proportional representation system, large voting district), (c)USA(Majoritarian electoral system,small voting district), (d)Australia(Majoritarian electoral system,small voting district), (e)Canada(Majoritarian electoral system,small voting district) and, (f)Taiwan(Proportional representation system, single non-transferable vote and medium voting district).}

\end{figure}

In this paper, we will show that although different 
countries adopt different electoral systems, the results of their
vote distributions can indeed be explained by two competing 
effects, viz., the Majority Effect and the Media Effect.  We will 
demonstrate this by mimicking these two effects in a simple growth
model and show that one can reproduce the distribution of votes in
different countries which adopt different electoral systems.  Using
models from physics to study topics in political science is common
nowadays.  For example, one can study political districting problem
by using the well-known Potts model in physics\cite{chou}.  

To facilitate our discussion, let us first define what we mean by
Majority and Media Effects.  By Majority Effect, we here mean
that the decision of a person to vote for a particular candidate is
mainly influenced by the portion of his neighbors and friends who 
want to vote for that particular candidate.  The larger the portion is, 
the more likely that the person will vote for this particular
candidate.  Hence we call it the Majority Effect.  It is easy to imagine 
that most of the less known candidates in fact obtain their votes 
through this process.  On the other
hand, a person can also obtain information of a particular
candidate through other sources such as the mass media.  For example, 
in the 2008 USA Presidential Election, most Americans obtained their 
information about Obama and McCain through mass media such as TV, 
internet, etc, much less from their friends and neighbors.  
In this way, we can say that people obtain 
their information about a certain candidate through some long range
interactions.  We categorize these long range interactions 
as the Media Effect.  We believe that the two effects indeed
compete against each other which results in different patterns of  vote 
distributions as observed in different countries.  

\section{Model}

We will investigate the two effects one by one and then 
combine the two effects in our growth model to understand how they
compete against each other.  
Let us first consider only the Majority Effect.  To make things
simple, we now put all voters on a regular lattice.  Let us first
put them on a one-dimensional regular lattice (ring 
if we impose periodic boundary condition).  Each of the voters will 
then take a site on this lattice.  In this case, each voter will 
only be affected by his neighbors whom they will vote for.  
To begin with, let us assume that there are $N$ voters (lattice sites) 
and $M$ candidates.  The candidates will act as the seeds on the 
lattice and they are randomly distributed on the lattice.  
In the first time step, the candidates start to 
convince his neighbors (neighboring sites) to vote for him.  
In the next time step, his neighbors (neighboring sites) 
will convince their neighbors (neighboring sites) to vote for the
same candidate.  When an undecided voter has decided voters on both 
sides, his vote will be determined by randomly picking one of his
neighbor's decision.  The process will stop when all the voters have
decided whom they will vote for.  This process is indeed equivalent to a 
simple one-dimensional growth process with $M$ seeds to begin with.  The 
distribution of $M$ random seeds (candidates) should follow an exponential 
distribution $e^{-(x/x_0)}$ where $x$ is the distance between 
two neighboring seeds and $x_0 = 1/M$, the average lattice distance
between two neighboring seeds.  We here normalize $x$ to be from 0 to 1.
At the end of the process, there will be a distribution of votes from all the
seeds(candidates).  The form of the distribution of
votes in this simple case can be obtained analytically.  The argument 
is as follows.  In order for a candidate to get a portion of votes
$x$, he has to have two nearest neighboring candidates to be of a distance $2x$
apart.  The probability to have two neighboring candidates of distance $2x$ apart
is $e^{-(2 x/x_0)}$.  The seed can be anywhere between these
two neighboring candidates.  Integrating out all possibilities will give a factor
proportional to $x$.  The vote distribution in this simple
case is therefore equal to $a x e^{-(2 x/x_0)}$, where $a$ is the
normalization factor.  $a$ can be fixed by integrating over $x$ from 
0 to 1/2, since the largest separation between two sites with a seed
in between should be less than 1.  Notice that there are no 
free parameters in this simple one-dimensional case.

One can easily extend this analysis to higher dimensional cases.
For a certain number of seeds randomly distributed on a two-dimensional
surface, their distribution takes the form $r e^{-\rho \pi r^2}$,
where $r$ is the distance measured from a specified seed and $\rho$ is the
density of seeds in the two-dimensional plane.  In a similar fashion,
the vote distribution on the two-dimensional square lattice would take a form
$a x^b e^{-c x}$, where $a, b, c$ are constants and $x$
is the portion of votes that a certain seed will get.  Figure 2 is an illustration of 
simulations of both the one- and two-dimensional lattices with only Majority Effect 
and with $N=1,000,000$ and $M=1,000$.  The curves from theoretical
arguments are also included for comparison.  In the two-dimensional case, the
theoretical fit gives $a = 6.19 \times 10^9, b = 0.012$ and $c = 2.48$ respectively.  

\begin{figure}
\includegraphics{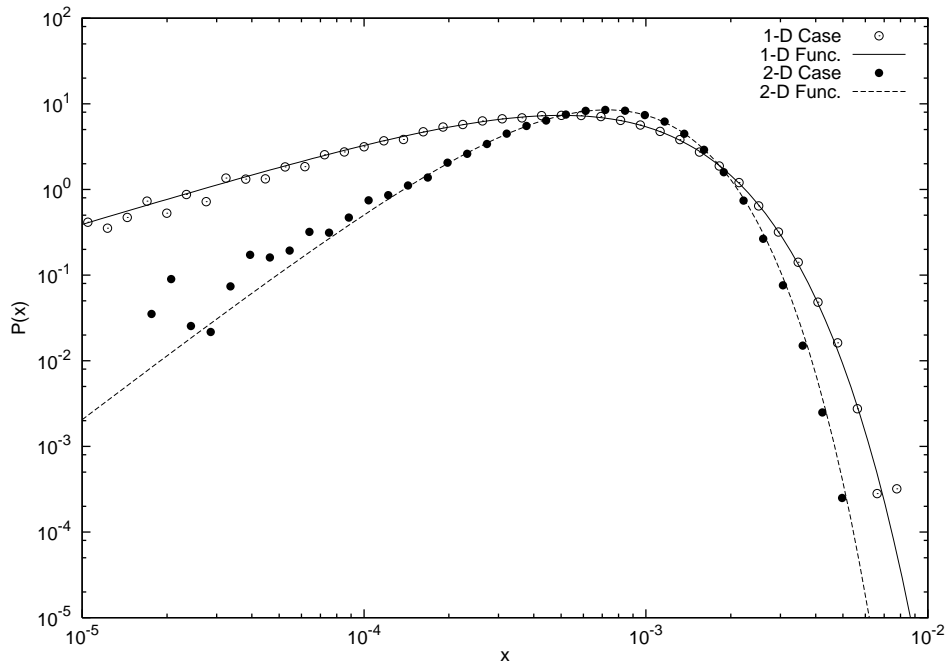}
\caption{\label{fig:2} Simulations of both the one- and two-dimensional lattices with 
only Majority Effect.  Here, $N=1,000,000$ and $M=1,000$.}
\end{figure}

The distribution of votes for the Majority Effect changes significantly if
one introduces an extra factor which takes the form

\begin{equation}
  \{ \frac{N_i}{\sum^M_{i=1} N_i}  \}^\beta  ,
\end{equation}

\noindent
where $i$ refers to the $i^{\texttt{th}}$ candidate, $N_i$ is the total number of
voters who have decided to vote for the this candidate and 
$\sum^M_{i=1} N_i$ is the total number of voters who have already decided
to vote for one of the $M$ candidates.  $\beta$ here is a free parameter.
The introduction of the factor in Eq.(1) has the following meaning.  
A decided voter would want to convince
his neighbors and friends to vote for that particular candidate.  However,
the probability that he could succeed would also depend on how well the 
candidate has been doing so far.  One can conceive that if the candidate had 
been doing very poorly, the probability that the decided voter could successfully 
convince his neighbors and friends to vote for the candidate would be low.  
On the other hand, if the candidate had been doing very well, it could have 
been much easier for the decided voter to convince his neighbors and friends 
to vote for the candidate.  The case of $\beta = 0$ is the
extreme case which corresponds to what we have discussed above.  This is 
the case when a decided voter can always convince his neighbors to 
vote for the candidate that he decided to vote for, regardless of how the 
candidate has been doing so far.  Figure 3 is an illustration of how
the distribution of votes will change as we change $\beta$.  As $\beta$
increases, the tails on both ends of the distribution will start to move up until 
it displays a power law like distribution when $\beta \approx 0.7$.  

\begin{figure}
\includegraphics{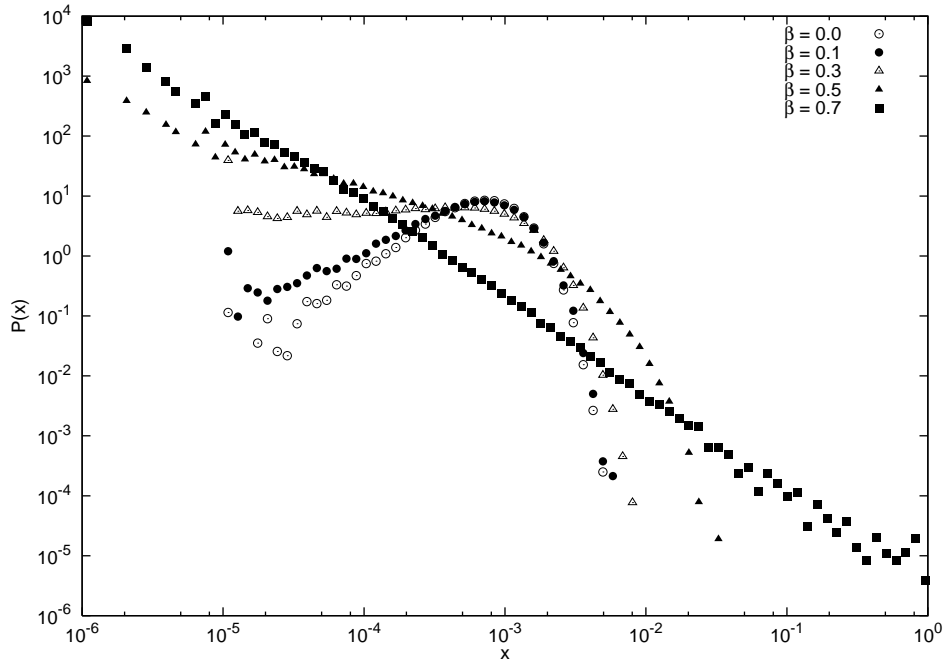}
\caption{\label{fig:3} Vote distributions with only Majority Effect on a two-dimensional 
square lattice as $\beta$ in Eq. (1) changes.}
\end{figure}

As mentioned above, a few of the candidates can obtain a large portion of 
votes largely by Media Effect in an election.  We investigate this effect again 
on a lattice.  We here again assume that there are $M$ candidates, each 
candidate can put $L$ random seeds on the lattice with a total number of 
$N$ sites.  Since the $M \times L$ seeds are randomly put on the lattice, one 
would then expect the distribution to follow a Gaussian distribution.  Figure 4 
shows the result of the distribution of votes with $M= 4$ candidates on a two 
dimensional square lattice with a total of $N = 500 \times 500$ sites.  We
have tried two different situations here.  In Figure 4(a), $L=512$ and the 
random seeds are all put on the lattice in the very beginning.  We let the
system evolve similar to the case of the Majority Effect as described 
above.  One can see that the curve can be fitted well by a Gaussian distribution 
with its peak around 0.25 as expected.  In Figure 4(b), we start with only one 
seed for each of the $M$ candidates.  In each time step, we give a probability 
$\alpha$ that one of the undecided sites will become the seed of one of the $M$ 
candidates.  Again, the curve can be described by a Gaussian distribution with its 
peak around 0.25.  The results in both cases agree well with our expectation.  
From the above analysis, we therefore propose that the function that describes 
the various effects on vote distribution in elections would take the general form

\begin{equation}
a x^b \exp({-(\frac{x-c}{\sigma})^\gamma }),
\end{equation}

\noindent
where $a, b, c, \gamma$ and $\sigma$ are all constants and with the vote
distributions of the two effects studied in the above as particular cases.

\begin{figure}
\includegraphics{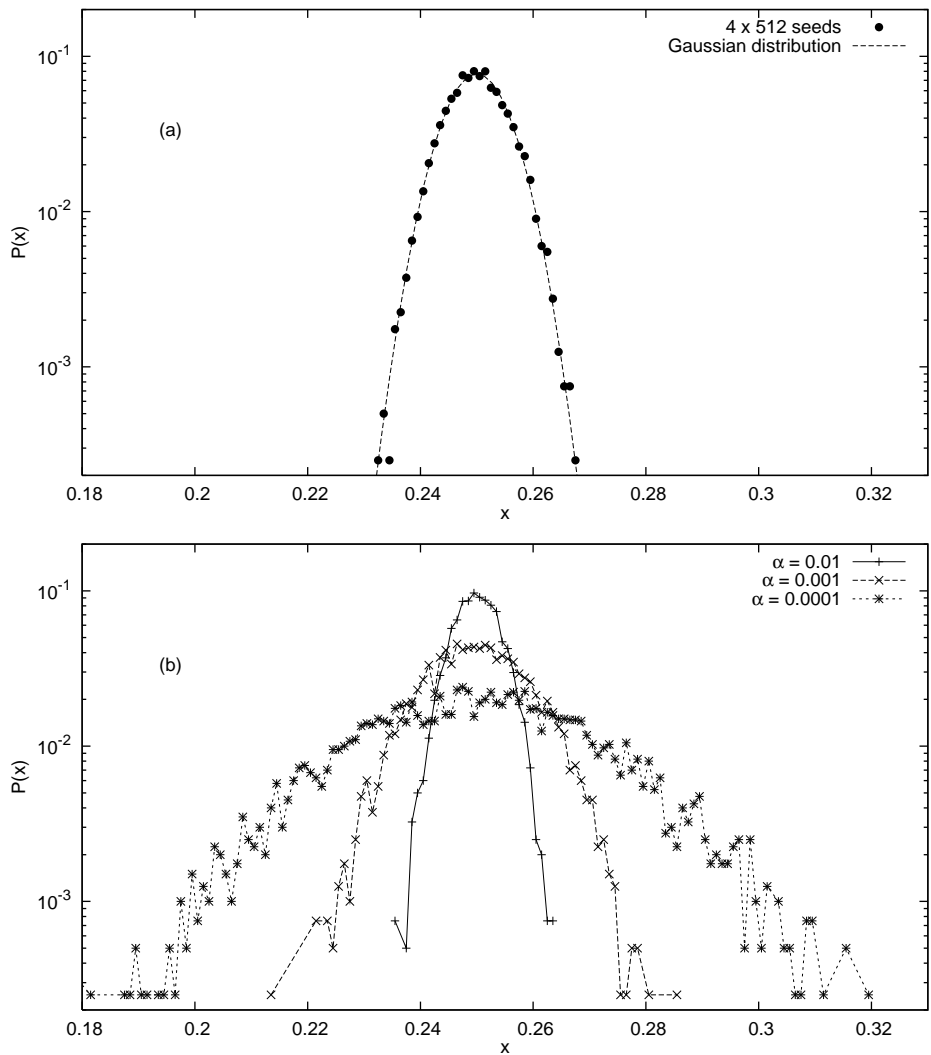}
\caption{\label{fig:4} Vote distributions with only Media Effect.  The number of candidates 
$M=4$ and each has $L=512$ random seeds on a square lattice with 250,000 sites.  
(a) All the random seeds are put on the lattice at the beginning and, (b) the number of seeds
grow as time evolves. }
\end{figure}

\section{Results}

We now include both effects in our growth model and investigate how they compete
against each other.  We again use the two-dimensional square lattice as 
an example.  To begin with, we have a number of candidates $M$ and 
a lattice with $N$ lattice sites.  Of the $M$ candidates, $M_1$ of them acquire 
Media Effect while the rest can only obtain their votes through Majority 
Effect.  The rules that mimic the two effects have been introduced above so
we will not repeat here.  Figure 5 are the distributions of votes with both 
effects included in the model.  Figures 5(a) - (f) are the fits using both our simulations 
and theoretical fit of the vote distributions corresponding to Figure 1(a)-(f).  
In each of the cases, we fit the real data using both simulations from our growth
model and the theoretical function of Eq. (2).  For example, Figure 5(a) is 
a fit of the 2006 election in Brazil.  The solid triangles are from real data while the circles
are the simulated result of our growth model with $M = 300$ and $N = 1,000,000 $.  
We use $\alpha = 0, \beta = 0.65$ in the simulation.  The solid line
is a theoretical fit using our function above with $a=9.94 \times 10^{19}, b=1.57, c=0, 
\sigma=2.38 \times 10^{-20}$ and $\gamma=0.089$.   Table 1 lists the values of 
$\alpha$ and $\beta$ that we use in order to fit the vote distributions in different 
countries.  We should also mention here that in the simulation, we use the same number
of voters (sites) corresponding to the number of voters in the election held by that country.
Therefore, even with the same set of $\alpha$ and $\beta$, the result of the simulation 
could be different.  An example is the simulations of Australia and Taiwan.  Even with 
the same values of $\alpha$ and $\beta$, the simulated results look somewhat different.
Theoretical fits for other 
countries can be done in a similar way but a sum of two functions similar to Eq. (2) 
is needed for Figure 5(c)-(f).   The peaks on the right of the distributions in Figure 5(c)-(f)
refer to the group of candidates who obtain their votes through Media Effect.  One
could therefore study the relative importance of the two effects in different 
electoral systems within our simple growth model which would have real applications
in elections.  

\begin{figure*}
\includegraphics[width=1.0\linewidth]{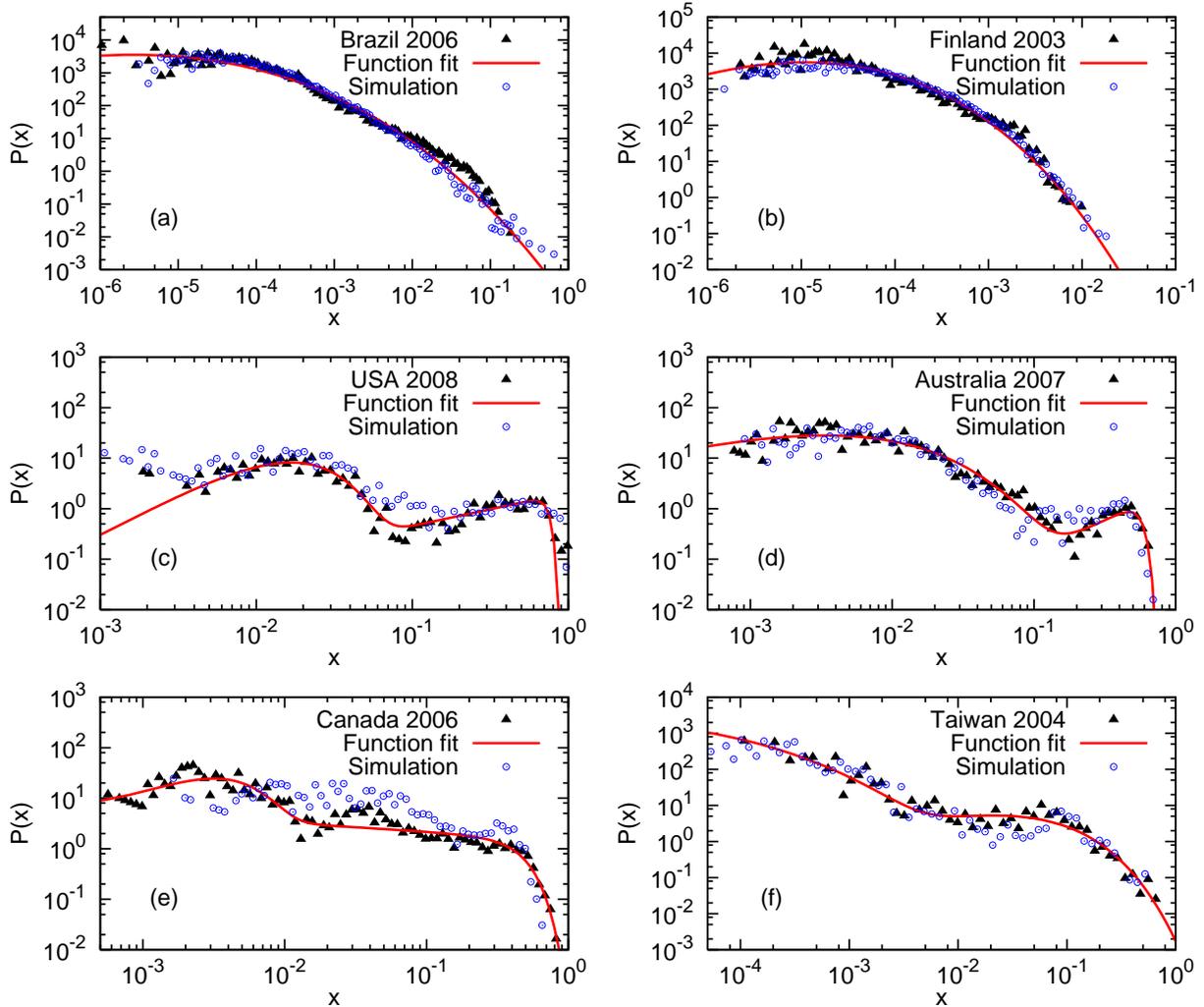}
\caption{\label{fig:5} Vote distributions from (a)Brazil, (b)Finland, (c)USA, (d)Australia, (e)Canada and, (f)Taiwan.}
\end{figure*}

\begin{table}
\caption{\label{tabone} Parameters of Media Effect ($\alpha$) and Majority Effect ($\beta$) in Figure 5. } 

{\begin{tabular}{@{}*{9}{l}}
                            
         && Brazil & Finland & USA    & Australia & Canada & Taiwan  \\ \hline 

$\alpha$ && 0.0    & 0.0     & 1.0E-6 & 1.0E-5    & 1.0E-5 &  1.0E-5 \\ \hline
$\beta$  && 0.65   & 0.6     & 0.35   & 0.6       & 0.55   &  0.6    \\ \hline

\end{tabular} }

\end{table}

\section{Discussion}

The fact that there are two different groups of candidates in an election might
be understood in the language of networks.  The candidates who only 
obtain votes through Majority Effect can be conceived as people who communicate 
with others through a regular network with nearest neighbor interactions and 
could only affect others through some kind of diffusion process which is rather slow.
On the other hand, candidates who obtain votes through Media Effect are more likely
connected to networks similar to small world or scale free networks in which
case they can affect voters from distance away.

In summary, we have identified two dominant effects in elections and
mimic them in a simple growth model.  We put the growth 
model on a two-dimensional square lattice and test it against the available data 
from elections in various countries.  By adjusting the two parameters $\alpha$ 
and $\beta$ in the model, we are able to fit the vote distributions of 
elections in all the countries that we have studied.  We also give a theoretical argument 
of the general form of the function it can take and fit it with data.  Other 
effects as well as lattices with higher dimensions
and/or different structures can be used and their effects on vote distributions
will be investigated in the future.

\end{document}